\documentclass[floats, aps]{revtex4}
\usepackage{amsmath}
\usepackage{graphicx}
\usepackage{epsfig}

\begin{document}
\title { Transverse Symmetry Transformations and the Quark-Gluon Vertex Function in QCD }

\author{Han-xin He $^{a,b,*}$ \footnotetext{$^*$ E-mail address: hxhe@ciae.ac.cn}}
\affiliation{ $^{a}$ China Institute of Atomic Energy, P.O.Box
275(18), Beijing 102413, China \\
$^{b}$ Department of Modern Physics, University of Science and
Technology of China, Hefei 230026, China }

\begin{abstract}

The transverse symmetry transformations associated with the normal
symmetry transformations in gauge theories are introduced, which
at first are used to reproduce the transverse Ward-Takahashi
identities in the Abelian theory QED. Then the transverse symmetry
transformations associated with the BRST symmetry and chiral
transformations in the non-Abelian theory QCD are used to derive
the transverse Slavnov-Taylor identities for the vector and
axial-vector quark-gluon vertices, respectively. Based on the set
of normal and transverse Slavnov-Taylor identities, an expression
of the quark-gluon vertex function is derived, which describes the
constraints on the structure of the quark-gluon vertex imposed
from the underlying gauge symmetry of QCD alone. Its role in the
study of the Dyson-Schwinger equation for the quark propagator in
QCD is discussed.


\end{abstract}
\maketitle

\section{Introduction}

Gauge symmetry imposes powerful constraints on the basic vertex
functions of gauge theories, referred to as the Ward-Takahashi(WT)
\cite{ward} or the Slavnov-Taylor(ST) identities \cite{slav}. They
play an essential role in demonstrating the renormalizability of
gauge theories and are also important in the nonperturbative
studies of gauge theories by using the Dyson-Schwinger
equations(DSEs)\cite{robe}. In these aspects, the knowledge of the
structure of the quark-gluon vertex is essential for understanding
the dynamics of quark confinement and chiral symmetry breaking and
also plays a key role in bridging the color quarks and gluons and
their colorless bound states (hadrons) \cite{robe,alk1,skul,munc}.

The quark-gluon vertex $\Gamma^{a\nu}(p_1,p_2)$ together with the
quark propagator $S_F(p_1)$ and gluon propagator $D_{\mu\nu}(q)$
enter as the vital ingredients in the DSE for the quark
propagator. The kernel of the quark DSE is dominated by
$D_{\mu\nu}(q)\Gamma^{a\nu}(p_1,p_2)$ with $q=p_1-p_2$. Using the
general form of the gluon propagator in covariant gauge, this
kernel is written as $D_{\mu\nu}(q)\Gamma^{a\nu}(p_1,p_2) =[ -
\frac{\xi}{q^2}\frac{q_{\mu}q_{\nu}}{q^2}-(g_{\mu\nu}-
\frac{q_{\mu}q_{\nu}}{q^2})\frac{Z_g}{q^2}]\Gamma^{a\nu}(p_1,p_2)$
with $Z_g$ being the gluon dressing function and $\xi$  being the
covariant gauge parameter, which can be expressed as
$D_{\mu\nu}(q)\Gamma^{a\nu}(p_1,p_2) = -
\frac{\xi}{q^2}q^{-2}q_{\mu} [q^{\nu} \Gamma
^a_{\nu}(p_1,p_2)]-\frac{Z_g}{q^2}q^{-2}iq^{\nu}[ iq_{\mu}\Gamma
^a_{\nu}(p_1,p_2)-iq_{\nu}\Gamma ^a_{\mu}(p_1,p_2)]$. This kernel
is clearly separated into the contributions from longitudinal and
transverse parts of the quark-gluon vertex. Notice that in Landau
gauge with $\xi=0$ the contribution from the longitudinal part of
the vertex to this kernel will disappear and the transverse part
of the vertex will dominate this kernel and then the DSE for the
quark propagator. The normal ST identity for the quark-gluon
vertex determines the longitudinal part of the vertex, but leaving
the transverse part unconstrained. It therefore appears highly
desirable to determine constraints on the transverse part of the
quark-gluon vertex from the gauge symmetry--the transverse ST
identities. It is known that the ST identity for the quark-gluon
vertex can be derived \cite{eich} by using the BRST (
Becchi-Rouet-Stora-Tyutin) symmetry \cite{becc}. Natural question
is then if there exists a kind of symmetry transformation which
enables to derive the constraint relation for the transverse part
of a basic vertex in gauge theories? Can it be used to derive the
complete constraints on the structure of the quark-gluon vertex ?
This paper provides the answer to them.

In this paper, the transverse symmetry transformations associated
with the normal symmetry transformations in gauge theories are
proposed, which are defined by the infinitesimal Lorentz
transformation for the normal symmetry transformations. At first,
we use the transverse symmetry transformations associated with the
normal symmetry transformations and chiral transformations in
Abelian theory QED in terms of the path-integral approach to
re-produce the transverse WT relations that obtained previously
based on the canonical field theory approach\cite{he1,he2}. Then
we show how the transverse symmetry transformations associated
with the BRST transformations and chiral transformations in QCD
can be used to derive the transverse Slavnov-Taylor identities for
the vector and axial-vector quark-gluon vertices, respectively, by
the path-integral approach. Based on the set of the normal and
transverse ST identities for the vector and the axial-vector
quark-gluon vertices, we further derive the quark-gluon vertex
function involving both longitudinal and transverse parts of the
vertex, which describes the constraints on the quark-gluon vertex
structure imposed from the gauge symmetry of QCD alone.

This paper is organized as follows. In Sec.II, the definition of
the transverse symmetry transformation associated with the normal
symmetry transformation is introduced. According to the
definition, in Sec.III we write the detailed representations for
the transverse symmetry transformations associated with the normal
and chiral transformations in QED, respectively, which are used to
re-derive the transverse WT relations for the vector and
axial-vector vertices in QED by the path-integral approach. Then
in Sec.IV we give the transverse symmetry transformations
associated with the BRST symmetry and the chiral transformations
in QCD, and then derive the transverse ST identities for the
vector and axial-vector quark-gluon vertices, respectively. By
using the set of the normal and transverse ST identities for the
vector and axial-vector quark-gluon vertices, in Sec.V we derive
the quark-gluon vertex function in QCD. The conclusions and
discussions are given in Sec.VI.

\section{Transverse Symmetry Transformations in Gauge Theories}

Let us begin with introducing the definition and representation of
the transverse symmetry transformations associated with normal
symmetry transformations in gauge theories. Consider an
infinitesimal symmetry transformation
\begin{equation}
\phi^a(x) \longrightarrow \phi^a(x) + \delta \phi^a(x).
\end{equation}
Now we introduce corresponding infinitesimal transverse symmetry
transformation
\begin{equation}
\phi^a(x) \longrightarrow  \phi^a(x) + \delta_{T} \phi^a(x),
\end{equation}
where $\delta_{T} \phi^a(x)$ is defined by the infinitesimal
Lorentz transformation for the infinitesimal symmetry
transformation $\delta \phi^a(x)$ in (1):
\begin{equation}
 \delta_{T} \phi^a(x) =
\delta_{Lorentz}(\delta \phi^a(x))
 = - \frac{i}{2}\epsilon^{\mu\nu}
 S_{\mu\nu}^{(\delta \phi^a)}(\delta \phi^a(x)).
\end{equation}
Here $S_{\mu\nu}^{(\delta \phi^a)}$ denotes the generator of the
intrinsic part for the infinitesimal Lorentz transformation for
$\delta \phi^a(x)$, where $\phi^a$ may be the spinor, vector or
scalar field, and $\delta \phi^a(x)$ may be composed of these
fields. For the spinor field,
\begin{equation}
 S_{\mu\nu}^{(spinor)}=\frac{1}{2}\sigma_{\mu\nu},
\end{equation}
where $\sigma_{\mu\nu}=\frac{i}{2}[\gamma_{\mu},\gamma_{\nu}]$;
for the vector field,
\begin{equation}
 (S_{\mu\nu}^{(vector)})^{\alpha}_{\beta}=i( \delta^{\alpha}_{\mu}g_{\nu\beta}-
 \delta^{\alpha}_{\nu}g_{\mu\beta} );
\end{equation}
and for the scalar field,
\begin{equation}
 S_{\mu\nu}^{(scalar)}=0.
\end{equation}

 The physical picture of the transformation (2), with the
definition (3), is clear: While the transformation (1) defines a
symmetry transformation where the change of variable is along the
symmetry direction, the transformation (2) transforms the original
symmetry direction, by the infinitesimal Lorentz transformation
(3), to its transverse direction. This is why we call the
transformation (2) with (3) as the transverse symmetry
transformation associated with the symmetry transformation (1),
explaining why the transverse WT(ST) identity for the vertex,
derived in terms of the corresponding transverse symmetry
transformation, can constrain the transverse part of the vertex.

In the following Sec.III and IV, we will use the definition (3) to
give the explicit representations for the transverse symmetry
transformations in the Abelian gauge theory QED and in the
non-Abelian gauge theory QCD, respectively, and then derive the
corresponding transverse WT identities in QED and transverse ST
identities in QCD.

\section{Transverse Ward-Takahashi Relations in QED}

The Ward-Takahashi(WT) identities are well-known to derive by
using the canonical field theory approach or the path-integral
approach. The transverse WT identities have been derived already
in the canonical field theory approach\cite{he1,he2}. As a check
to the transverse symmetry transformation approach, in this
section we provide the re-derivations of the transverse WT
identities in terms of transverse symmetry transformations by
using the path-integral approach. The procedure and results will
be helpful also to understand the derivations of transverse ST
identities in QCD given in next section.

In the path-integral approach, the origin of the WT(ST) identities
lies in the gauge invariance of the generating functional of
QED(QCD). If making the infinitesimal gauge transformations of the
fields in QED: $\psi(x) \longrightarrow \psi'(x),\\\
 \bar{\psi}(x) \longrightarrow  \bar{\psi}'(x)$
and $A_{\mu}(x)\longrightarrow A_{\mu}'(x)$, the generating
functional of QED remains the same, which leads to the following
identity for the functional integral over two fermion fields:
\begin{equation}
\int D[\psi,\bar{\psi},A]e^{i\int d^4xL_{QED}[\psi,\bar{\psi},A]}
\psi(x_1)\bar{\psi}(x_2)= \int D[\psi',\bar{\psi}',A']e^{i\int
d^4xL_{QED}[\psi',\bar{\psi'},A']} \psi'(x_1)\bar{\psi}'(x_2).
\end{equation}
Further considering if the measure of functional integration is
invariant or not, one then can derive the relative WT relation in
QED from the identity (7). The ST relations in QCD can be derived
by the parallel way.

The infinitesimal gauge transformations of the fields in QED can
be written as
\begin{equation}
\psi(x) \longrightarrow \psi'(x)= \psi(x) + ig\alpha(x)\psi(x),\\\
 \bar{\psi}(x) \longrightarrow  \bar{\psi}'(x) =\bar{\psi}(x) -
ig\alpha(x)\bar{\psi}(x),
\end{equation}
and $A_{\mu}\longrightarrow A_{\mu}$ or $A_{\mu}\longrightarrow
A_{\mu}-\partial_{\mu}\alpha(x)$. Here
$\delta\psi(x)=ig\alpha(x)\psi(x),
\delta\bar{\psi}(x)=-ig\alpha(x)\bar{\psi}(x)$, where $g = -e$ and
$\alpha(x)$ is an infinitesimal real parameter. Using the identity
(7) and the fact that the measure of functional integration is
invariant under the gauge transformations given by Eq.(8), it is
easy to write the WT identity for the fermion-boson vertex in
coordinate space\cite{ward,pesk}:
\begin{equation}
\partial^{x}_{\mu} \langle 0|Tj^{\mu}(x) \psi (x_{1})
\bar{\psi}(x_{2})|0\rangle = \langle 0|T \psi (x_{1}) \bar{\psi}
(x_{2})|0\rangle [\delta^{4} (x - x_{2}) - \delta^{4}(x - x_{1})]
\end{equation}
with $j^{\mu}(x)=\bar{\psi}(x)\gamma ^{\mu }\psi (x)$, which in
momentum space gives the familiar expression
\begin{equation}
q_{\mu} \Gamma_V^{\mu} (p_{1}, p_{2}) = S_{F}^{-1} (p_{1}) -
S_{F}^{-1} (p_{2}) ,
\end{equation}
where $q=p_1 - p_2$, $S_F(p)$ is the full fermion propagator, and
$\Gamma_V^{\mu}$ is defined by
\begin{eqnarray}
& &\int d^{4}x d^{4}x_{1} d^{4}x_{2} e^{i(p_{1}\cdot  x_{1} -
p_{2}\cdot  x_{2} - q\cdot  x)} \langle 0|T j^{\mu}(x)\psi(x_{1})
\bar{\psi}(x_{2}) |0\rangle \nonumber \\
&=& (2 \pi)^{4} \delta^{4}(p_{1} - p_{2} - q) iS_{F}(p_{1})
\Gamma^{\mu}_V iS_{F}(p_{2}) .
\end{eqnarray}
The transverse WT identities can be derive by the parallel
procedure.

\subsection{Transverse WT identity for the fermion-boson vertex from transverse
symmetry transformations}

The infinitesimal transverse symmetry transformations associated
with the symmetry transformations given by Eq.(8) can be written
by the definition (3):
\begin{equation}
\delta_T\psi(x)=\frac{1}{4}g\alpha(x)\epsilon^{\mu\nu}\sigma_{\mu\nu}\psi(x),\
\ \
\delta_T\bar{\psi}(x)=\frac{1}{4}g\alpha(x)\epsilon^{\mu\nu}\bar{\psi}(x)\sigma_{\mu\nu},
\end{equation}
without the corresponding transformation term for $A_{\mu}$. Under
such infinitesimal transverse symmetry transformations, the QED
lagrangian transforms according to $L_{QED} \longrightarrow
L_{QED} + \delta_T L_{QED}$, where
\begin{eqnarray}
\delta_TL_{QED}&=&\frac{i}{4}g\alpha(x)\epsilon^{\mu\nu}\bar{\psi}(x)
S_{\lambda\mu \nu }(\overrightarrow{\partial}^\lambda_x-
\overleftarrow{\partial}^\lambda _x)\psi(x) +
\frac{1}{2}g^2\alpha(x)\epsilon^{\mu\nu}\bar{\psi}(x)S_{\lambda
\mu \nu}A_{\lambda}\psi(x)\nonumber \\
& & -
\frac{1}{2}mg\alpha(x)\epsilon^{\mu\nu}\bar{\psi}(x)\sigma_{\mu\nu}\psi(x)
-\frac{1}{4}g\epsilon^{\mu\nu}(j_{\nu}(x)\partial_{\mu}\alpha(x)
-j_{\mu}(x)\partial_{\nu}\alpha(x)).
\end{eqnarray}
Here $S_{\lambda\mu
\nu}=\frac{1}{2}\{\gamma_{\lambda},\sigma_{\mu\nu}\}=-\varepsilon
_{\lambda \mu \nu \rho }\gamma^{\rho}\gamma_5$. In order to relate
each term on the right-hand side of Eq.(13) to a definite Green's
function, we need to move the derivative operators in the first
term on the right-hand side of Eq.(13) out of the current operator
$\bar{\psi}(x)S_{\lambda \mu \nu}\psi(x)$. For this purpose, we
first write $\bar{\psi}(x)S_{\lambda \mu \nu}\psi(x)$ as
$\bar{\psi}(x^{\prime})S_{\lambda \mu \nu}\psi(x)$, and then at
the end of the calculation we take $x^{\prime }\rightarrow x$. But
the new operator including the nonlocal current is not gauge
invariant. To recover the gauge invariant expression, we use the
standard procedure by introducing the line integral, i.e. the
Wilson line $U_P (x^{\prime },x)=P\exp (ig\int_x^{x^{\prime
}}dy^\rho A_\rho (y))$ between $\bar{\psi}(x^{\prime})$ and
$\psi(x)$( also see\cite{he1} for discussion ). We thus obtain
\begin{eqnarray}
\delta_TL_{QED}&=&\frac{i}{4}g\alpha(x){\lim_{x^{\prime
}\rightarrow x}}\epsilon^{\mu\nu}(\partial ^\lambda _x-\partial
^\lambda_{x^{\prime }})\bar{\psi}(x^{\prime})S_{\lambda\mu \nu }
U_P (x^{\prime},x)\psi(x) \nonumber \\
& & -\frac{1}{2}
mg\alpha(x)\epsilon^{\mu\nu}\bar{\psi}(x)\sigma_{\mu\nu}\psi(x)
-\frac{1}{4}g\epsilon^{\mu\nu}(j_{\nu}(x)\partial_{\mu}\alpha(x)
-j_{\mu}(x)\partial_{\nu}\alpha(x)).
\end{eqnarray}
The form of Eq.(14) insures that each term on the right-hand side
of equation can be related to a definite Green's function. Since
the measure of functional integral is invariant under the
transformations given by Eq.(12), such transformations then lead
to the following identity from Eq.(7) for the functional integral
over two fermion fields
\begin{equation}
0=\int D[\bar{\psi},\psi, A] e^{i\int d^4xL_{QED}}\{ i\int
d^4x(\delta_T L_{QED})\psi(x_1)\overline{\psi}(x_2)
+\delta_T(\psi(x_1)\overline{\psi}(x_2))\}.
\end{equation}
 Substituting Eqs.(12) and (14) into Eq.(15) and
integrating the term involving $\partial\alpha$ by parts, and then
taking the coefficient of $\alpha$ and dividing the generating
functional $Z[J=0]$, we obtain the transverse WT identity for the
fermion-boson (vector) vertex in coordinate space:
\begin{eqnarray}
& &\partial _x^\mu \left\langle 0\left| Tj^\nu (x)\psi
(x_1)\bar{\psi}(x_2) \right| 0\right\rangle -\partial _x^\nu
\left\langle 0\left| Tj^\mu (x)
\psi (x_1)\bar{\psi}(x_2)\right| 0\right\rangle \nonumber \\
&=&i\sigma ^{\mu \nu }\left\langle 0\left| T\psi
(x_1)\bar{\psi}(x_2)\right| 0\right\rangle \delta
^4(x_1-x)+i\left\langle 0\left| T\psi (x_1)\bar{\psi}(x_2)\right|
0\right\rangle
\sigma ^{\mu \nu }\delta ^4(x_2-x)\nonumber \\
& &+2m\left\langle 0\left| T\bar{\psi}(x)\sigma ^{\mu \nu }\psi
(x)\psi (x_1)
\bar{\psi}(x_2)\right| 0\right\rangle \nonumber \\
& &+{\lim _{x^{\prime }\rightarrow x}}i(\partial _\lambda
^x-\partial _\lambda ^{x^{\prime }}) \varepsilon ^{\lambda \mu \nu
\rho }\left\langle 0\left| T\bar{\psi}(x^{\prime }) \gamma _\rho
\gamma _5U_P (x^{\prime },x)\psi (x)\psi (x_1)\bar{\psi}(x_2)
\right| 0\right\rangle .
\end{eqnarray}
The result is same as that obtained previously by the canonical
field theory approach\cite{he1}.

By carefully computing the Fourier transformation of Eq.(16), we
obtain the transverse WT identity for the fermion-boson vertex
function in momentum space:
\begin{eqnarray}
& &iq^\mu \Gamma _V^\nu (p_1,p_2)-iq^\nu \Gamma _V^\mu (p_1,p_2)\nonumber \\
&=&S_F^{-1}(p_1)\sigma ^{\mu \nu }+\sigma ^{\mu \nu }S_F^{-1}(p_2)
+ 2m\Gamma _T^{\mu \nu }(p_1,p_2)\nonumber \\
& &+(p_{1\lambda }+p_{2\lambda })\varepsilon ^{\lambda \mu \nu
\rho } \Gamma _{A\rho }(p_1,p_2) -\int \frac{d^{4}k}{(2
\pi)^{4}}2k_{\lambda} \varepsilon ^{\lambda \mu \nu \rho }\Gamma
_{A\rho }(p_1,p_2;k),
\end{eqnarray}
where the integral term involves $\Gamma_{A\rho}(p_1,p_2;k)$ with
the internal momentum $k$ of the gauge boson appearing in the
Wilson line. $\Gamma_{A\rho}(p_1,p_2;k)$ is defined by
\begin{eqnarray}
& &\int d^{4}x d^{4}x^{\prime}d^{4}x_{1} d^{4}x_{2}
e^{i(p_{1}\cdot x_{1} - p_{2}\cdot  x_{2} + (p_2-k)\cdot x -
(p_1-k)\cdot x^{\prime})} \langle 0|T \bar{\psi}(x^{\prime
})\gamma _\rho \gamma _5 U_P (x^{\prime },x)\psi (x)\psi
(x_1)\bar{\psi}(x_2) |0\rangle \nonumber \\
&=& (2 \pi)^{4} \delta^{4}(p_{1} - p_{2} - q) iS_{F}(p_{1})
\Gamma_{A\rho}(p_{1}, p_{2};k) iS_{F}(p_{2}) ,
\end{eqnarray}
where $q=(p_1-k)-(p_2-k)$. Using this definition, we can write the
explicit expression of $\Gamma_{A\rho}(p_{1}, p_{2};k)$ in the
perturbation theory order by order. For example, we have at
one-loop order
\begin{eqnarray}
& &\int \frac{d^{4}k}{(2 \pi)^{4}}2k_{\lambda}
\varepsilon ^{\lambda \mu \nu \rho }\Gamma _{A\rho }(p_1,p_2;k)\nonumber \\
&=& g^2\int \frac{d^{4}k}{(2 \pi)^{4}}2k_{\lambda} \varepsilon
^{\lambda \mu \nu \rho }\gamma^{\alpha} \frac{1}{{\makebox[-0.8
mm][l]{/}{p}}_1- {\makebox[-0.8
mm][l]{/}{k}}-m}\gamma^{\rho}\gamma_5 \frac{1}{{\makebox[-0.8
mm][l]{/}{p}}_2-{\makebox[-0.8 mm][l]{/}{k}}-m}\gamma^{\beta}
\frac{-i}{k^2}[g_{\alpha\beta}+(\xi-1)\frac{k_{\alpha}k_{\beta}}{k^2}]\nonumber \\
& &+g^2\int \frac{d^{4}k}{(2 \pi)^{4}}2 \varepsilon ^{\alpha \mu
\nu \rho }[\gamma^{\beta} \frac{1}{{\makebox[-0.8
mm][l]{/}{p}}_1-{\makebox[-0.8
mm][l]{/}{k}}-m}\gamma^{\rho}\gamma_5 + \gamma^{\rho}\gamma_5
\frac{1}{{\makebox[-0.8 mm][l]{/}{p}}_2-{\makebox[-0.8
mm][l]{/}{k}}-m}\gamma^{\beta}]
\frac{-i}{k^2}[g_{\alpha\beta}+(\xi-1)\frac{k_{\alpha}k_{\beta}}{k^2}]
,
\end{eqnarray}
where ${\makebox[-0.8 mm][l]{/}{k}}=\gamma_{\mu}k^{\mu}$, and
$\xi$ is the covariant gauge parameter. The last two terms in the
right-hand side of Eq.(19) are the one-loop self-energy
contributions accompanying the vertex correction. It hence has
been checked that the transverse WT identity for the fermion-boson
vertex, Eq.(17), holds by the explicit computations of terms in
this transverse WT relation to one-loop
order\cite{he3,he4,penn,he5}. In Ref.\cite{penn} Pennington and
Williams also discussed the possibility to construct consistent
non-perturbative Feynman rules by using the transverse WT identity
(17).

\subsection{Transverse WT identity for the axial-vector vertex from
the transverse chiral transformations}

The infinitesimal chiral transformations in QED are known as
\begin{equation}
\psi(x) \longrightarrow  \psi(x) + ig\alpha(x)\gamma^5\psi(x),\\\
 \bar{\psi}(x) \longrightarrow  \bar{\psi}(x) +
ig\alpha(x)\bar{\psi}(x)\gamma^5,
\end{equation}
where $\delta^{(5)}\psi(x)=ig\alpha(x)\gamma^5\psi(x)$,
$\delta^{(5)}\bar{\psi}(x)=ig\alpha(x)\bar{\psi}(x)\gamma^5$,
which lead to the axial-vector WT identity (i.e. the WT identity
for the axial-vector vertex ) in coordinate space :
\begin{eqnarray}
& &\partial^{x}_{\mu} \langle 0|Tj^{\mu}_5(x) \psi (x_{1})
\bar{\psi}(x_{2})|0\rangle \nonumber \\
&=& -\delta^{4}(x - x_{1})\gamma^5 \langle 0|T \psi (x_{1})
\bar{\psi} (x_{2})|0\rangle - \delta^{4}(x - x_{2})\langle 0|T
\psi (x_{1}) \bar{\psi} (x_{2})|0\rangle \gamma^5 \nonumber \\
& &-\frac{e^2}{16\pi^2}\varepsilon^{\mu\nu\lambda\rho}
 \langle 0|T \psi(x_{1})
\bar{\psi}(x_{2}) F_{\mu \nu}(x) F_{\lambda\rho }(x) |0\rangle
\end{eqnarray}
with $j^{\mu}_5(x)=\bar{\psi}(x)\gamma ^{\mu }\gamma_5\psi (x)$.
The last term in Eq.(21) denotes the contribution of the axial
anomaly\cite{adle}, which arises from the change of functional
integration measure under the chiral transformations in the
path-integral approach\cite{fuji}. The axial-vector WT identity in
momentum space is given by Fourier transforming (21):
\begin{equation}
q_\mu \Gamma _A^\mu (p_1,p_2)=S_F^{-1}(p_1)\gamma _5+\gamma
_5S_F^{-1}(p_2) + i \frac{g^2}{16\pi^2} F(p_1,p_2) ,
\end{equation}
where $F(p_1,p_2) $ denotes the contribution of the axial anomaly
in momentum space and is defined by
\begin{eqnarray}
& &\int d^{4}x d^{4}x_{1} d^{4}x_{2} e^{i(p_{1}\cdot  x_{1} -
p_{2}\cdot  x_{2} - q\cdot  x)} \langle 0|T \psi(x_{1})
\bar{\psi}(x_{2})\varepsilon^{\mu \nu \rho \sigma} F_{\mu \nu}(x)
F_{\rho \sigma}(x) |0\rangle \nonumber \\
&=& (2 \pi)^{4} \delta^{4}(p_{1} - p_{2} - q) iS_{F}(p_{1})
F(p_{1}, p_{2})iS_{F}(p_{2}) .
\end{eqnarray}

The infinitesimal transverse chiral-transformations associated
with the chiral transformations given by Eq.(20) can be obtained
by using the definition (3):
\begin{equation}
\delta^{(5)}_{T}\psi(x)=\frac{1}{4}g\alpha(x)\epsilon^{\mu\nu}\sigma_{\mu\nu}
\gamma^5\psi(x),\ \ \ \delta^{(5)}_{T}\bar{\psi}(x)= -
\frac{1}{4}g\alpha(x)\epsilon^{\mu\nu}\bar{\psi}(x)
\sigma_{\mu\nu}\gamma^5.
\end{equation}
Under such infinitesimal transverse chiral-transformations, the
QED lagrangian transforms according to $L_{QED} \longrightarrow
L_{QED} + \delta^{(5)}_{T} L_{QED}$, where
\begin{eqnarray}
\delta^{(5)}_{T}L_{QED} &=&\frac{i}{4}g\alpha(x){\lim_{x^{\prime
}\rightarrow x}}\epsilon^{\mu\nu}(\partial ^\lambda _x-\partial
^\lambda_{x^{\prime }})\bar{\psi}(x^{\prime})S_{\lambda\mu \nu
}\gamma_5U_P (x^{\prime},x)
\psi(x) \nonumber \\
& &
-\frac{1}{4}g\epsilon^{\mu\nu}(j_{5\nu}(x)\partial_{\mu}\alpha(x)
-j_{5\mu}(x)\partial_{\nu}\alpha(x))
\end{eqnarray}
with $j_{5\mu}(x)=\bar{\psi}(x)\gamma _{\mu} \gamma_{5} \psi (x)$.
The calculation shows that the measure of functional integration
is invariant under the transformations given by Eq.(24). Hence the
transverse chiral transformations lead to a similar identity as
given by Eq.(15) where $\delta_{T}$ is replaced with
$\delta^{(5)}_{T}$. Substituting Eqs.(24) and (25) into such
identity, we obtain the transverse WT identity for the
axial-vector vertex in coordinate space:
\begin{eqnarray}
& &\partial _x^\mu \left\langle 0\left| Tj_5^\nu (x)\psi
(x_1)\bar{\psi}(x_2) \right| 0\right\rangle -\partial _x^\nu
\left\langle 0\left| Tj_5^\mu (x) \psi (x_1)\bar{\psi}(x_2)\right|
0\right\rangle
\nonumber \\
&=&i\sigma ^{\mu \nu }\gamma _5\left\langle 0\left| T\psi
(x_1)\bar{\psi}(x_2) \right| 0\right\rangle \delta
^4(x_1-x)-i\left\langle 0\left| T\psi (x_1) \bar{\psi}(x_2)\right|
0\right\rangle \sigma ^{\mu \nu }\gamma _5
\delta ^4(x_2-x) \nonumber \\
& &+i\int d^{4}x^{\prime}\delta(x^{\prime} -x)(\partial_\lambda^x-
\partial _\lambda ^{x^{\prime }})\varepsilon ^{\lambda \mu \nu \rho }
\left\langle 0\left| T\bar{\psi}(x^{\prime })\gamma _\rho U_P
(x^{\prime },x) \psi (x)\psi (x_1)\bar{\psi}(x_2) \right|
0\right\rangle ,
\end{eqnarray}
which is also same as the result obtained by the canonical field
approach\cite{he2}. Note that there is no transverse axial-anomaly
\cite{he6,sun}, which is consistent with the fact that the measure
of functional integration is invariant under the transverse chiral
transformations.

The transverse axial-vector WT identity in momentum space is given
by Fourier transforming Eq.(26) and the result is
\begin{eqnarray}
& &iq^\mu \Gamma _A^\nu (p_1,p_2)-iq^\nu \Gamma _A^\mu (p_1,p_2)\nonumber \\
&=&S_F^{-1}(p_1)\sigma ^{\mu \nu }\gamma _5- \sigma ^{\mu \nu
}\gamma _5S_F^{-1}(p_2)\nonumber \\
& &+(p_{1\lambda }+p_{2\lambda })\varepsilon ^{\lambda \mu \nu
\rho } \Gamma _{V\rho }(p_1,p_2) -\int \frac{d^{4}k}{(2
\pi)^{4}}2k_{\lambda} \varepsilon ^{\lambda \mu \nu \rho }\Gamma
_{V\rho }(p_1,p_2;k),
\end{eqnarray}
where $\Gamma _{V\rho}(p_1,p_2;k)$ is defined by the Fourier
transformation of the last matrix element in Eq.(26), which can be
written from Eq.(18) by replacing $\gamma _\rho\gamma _5$ and
$\Gamma_{A\rho}$ with $\gamma _\rho$ and $\Gamma_{V\rho}$,
respectively.

Now there are normal WT identities, which impose the constraints
on longitudinal parts of the vector and axial-vector vertices
given by Eqs.(10) and (22), respectively, and transverse WT
relations given by Eqs.(17) and (27), which constrain the
transverse parts of these vertices. The full constraint relations
for the vector and axial-vector vertex functions imposed from the
gauge symmetry alone then can be derived in terms of this set of
longitudinal and transverse WT relations\cite{he7} in QED. It also
has been checked that, by the explicit computations, such a full
fermion-boson vertex in QED holds\cite{he7} to one-loop order .

\section{Transverse Slavnov-Taylor Relations in QCD}

Now we study how the transverse symmetry transformations
associated with the BRST symmetry and the chiral transformations
enable us to derive the constraint relations for the transverse
parts of the vector and axial-vector quark-gluon vertices in QCD.
To do these, we begin with the BRST transformations\cite{becc} in
QCD:
\begin{equation}
\delta\psi = ig\omega c_{a}t^{a}\psi, \ \ \ \delta\bar{\psi} =
-ig\bar{\psi}t^{a}\omega c_{a}, \ \ \ \delta A^{a}_{\mu} = \omega
D^{ab}_{\mu} c_{b},\ \ \ \delta c^{a} = -\frac{1}{2}g\omega
f^{abc}c_{b}c_{c}, \ \ \ \delta \bar{c}^{a} = \frac
{\omega}{\xi}\partial^{\mu} A_{\mu}^{a},
\end{equation}
where $\psi$, $A_{\mu}^{a}$ and $c^a$ denote the quark, gluon and
ghost fields, respectively, $t^a$ is the generator of $SU(N_c)$
with $f^{abc}$ being the corresponding structure constants,
$\omega$ is an infinitesimal Grassmann number, $\xi$ is the
covariant gauge parameter, $D_{\mu}=\partial_{\mu}-igt^aA^a_{\mu}$
and $D^{ab}_{\mu} c_{b}=\partial_{\mu}c^a - gf^{abc}A_{\mu}^cc_b$.
The QCD action
\begin{eqnarray}
\int d^{4}xL_{QCD}&=&\int d^{4}x \{\bar{\psi}(x)(i
\gamma^{\mu}D_{\mu} -m )\psi(x)-
\frac{1}{4}F^a_{\mu\nu}(x)F^{a\mu\nu}(x) \nonumber \\
& & - \frac{1}{2\xi}(\partial^{\mu}A_{\mu}(x))^2 -
\partial^{\mu}\bar{c}^a(x)D^{ab}_{\mu}c^b(x)\}
\end{eqnarray}
is invariant under the BRST transformations, which can be used to
derive some useful ST identities. The Slavnov-Taylor identity for
the quark-gluon vertex in coordinate space reads:
\begin{eqnarray}
& &\frac{1}{\xi}\langle 0| T\psi (x_1)\bar{\psi}(x_2)
\partial^{\mu}A_{\mu}^a(x)| 0\rangle \nonumber \\
&=& -igt^{b}\langle 0| T c^b(x_1)\psi(x_1)
\bar{\psi}(x_2)\bar{c}^{a}(x)| 0\rangle + ig\langle 0| T\psi
(x_1)\bar{\psi}(x_2)c^{b}(x_2)\bar{c}^a{x}(0)| 0\rangle t^{b}.
\end{eqnarray}
Defining
\begin{eqnarray}
& &\int d^{4}x d^{4}x_{1} d^{4}x_{2} e^{i(p_{1}\cdot  x_{1} -
p_{2}\cdot  x_{2} - q\cdot  x)} \langle 0|T \psi(x_{1})
\bar{\psi}(x_{2})A^a_{\mu}(x) |0\rangle \nonumber \\
&=& (2 \pi)^{4} \delta^{4}(p_{1} - p_{2} - q) iS_{F}(p_{1})
ig\Gamma^{b\nu}_V iS_{F}(p_{2})iD^{ba}_{\nu\mu}(q) ,
\end{eqnarray}
Fourier transforming Eq.(30) and using the ST identity
$q_{\mu}D^{\mu\nu}_{ab}=-\xi\delta_{ab}q^{\nu}/q^2$, one then
obtains the ST identity for the quark-gluon vertex $\Gamma
^{a\mu}_{V}$ in momentum space\cite{eich}
\begin{equation}
q_\mu \Gamma ^{a\mu}_{V}(p_1,p_2;q)
=[S_F^{-1}(p_1)(t^{a}-B^{a}_4(p_1,p_2)) -
(t^{a}-B^{a}_4(p_1,p_2))S_F^{-1}(p_2)]G(q^2),
\end{equation}
where $q=p_1-p_2$, $G(q^2)$ is the ghost dressing function
relating to the ghost propagator by
\begin{equation}
D_G^{ab}(q)= - \delta^{ab}G(q^2)/q^2,
\end{equation}
and $B^a_4(p_1,p_2)$ is the 4-point quark-ghost scattering kernel
defined by
\begin{eqnarray}
& &gt^a\int d^{4}x d^{4}x_{1} d^{4}x_{2} e^{i(p_{1}\cdot  x_{1} -
p_{2}\cdot  x_{2} - q\cdot  x)} \langle 0|T \psi(x_{1})
\bar{\psi}(x_{2})c^a(x_1)\bar{c}^b(x) |0\rangle \nonumber \\
&=& (2 \pi)^{4} \delta^{4}(p_{1}-p_{2}-q)
g(t^{a}-B^{a}_4(p_1,p_2)) iS_{F}(p_{2})iD^{ba}_G(q) .
\end{eqnarray}

\subsection{Transverse Slavnov-Taylor identity for the quark-gluon vertex}

The infinitesimal transverse symmetry transformations of the
fields associated with the BRST transformations given by Eq.(28)
can be written by using definition (3). The result is
\begin{equation}
\delta_{T}\psi = \frac{1}{4}g \epsilon^{\mu\nu}\omega
c_{a}t^{a}\sigma_{\mu\nu} \psi, \ \ \  \delta_{T}\bar{\psi} =
\frac{1}{4}g\epsilon^{\mu\nu}\bar{\psi}\sigma_{\mu\nu} t^{a}\omega
c_{a},\ \ \ \delta_{T} A^{a}_{\mu} = \omega
\epsilon^{\mu\nu}D_{\nu}^{ab}c_{b}, \ \ \  \delta_{T} c^{a} =
\delta_{T} \bar{c}^{a} = 0.
\end{equation}

Under such transverse symmetry transformations, the QCD action
transforms according to $\int d^4xL_{QCD} \longrightarrow \int
d^4xL_{QCD} + \int d^4x\delta_T L_{QCD}$, where $\int d^4x\delta_T
L_{QCD}$ includes the contributions from quark, gluon,
gauge-fixing and ghost parts:
\begin{eqnarray}
\int d^4x\delta_T L_{quark} &=& \int d^4x \frac{1}{2}\omega g
\epsilon^{\mu\nu}\{\frac{i}{4}{\lim _{x^{\prime }\rightarrow
x}}(\partial _\lambda ^x-\partial _\lambda ^{x^{\prime }})
\bar{\psi} (x^{\prime}) S_{\lambda \mu \nu } \{U_P(x^{\prime },x),
t^{a}\} \psi (x)c^a(x) \nonumber \\
& & -m\bar{\psi}(x)\sigma _{\mu \nu }t^{a}\psi (x)c^{a}(x)+
3\bar{\psi}(x)\gamma_{\mu} t^{a}\psi(x)\partial_{\nu}c_{a}(x)
-3gf_{abc}\bar{\psi}(x)\gamma_{\mu}
t^{c}\psi(x)A_{\nu}^bc^{a}(x)\},
\end{eqnarray}
\begin{equation}
\int d^4x\delta_T L_{gluon} = \omega \epsilon^{\mu\nu}\int d^4x \{
\partial_{\lambda} F^a_{\lambda\mu} D^{ab}_{\nu}c_{b}(x) -
gf_{abc}F^a_{\lambda\mu} A^b_{\lambda}D^{cd}_{\nu}c_{d}(x) \},
\end{equation}
\begin{equation}
\int d^4x\delta_T L_{GF} = -\frac{1}{\xi}\omega
\epsilon^{\mu\nu}\int d^4x
\partial_{\lambda} A^a_{\lambda}(x)\partial_{\mu}(D^{ab}_{\nu}c_{b}(x)),
\end{equation}
\begin{equation}
\int d^4x\delta_T L_{ghost} = \omega \epsilon^{\mu\nu}\int d^4x
gf_{acd}\partial_{\mu} \bar{c}^c(x)c^d(x)D^{ab}_{\nu}c_{b}(x).
\end{equation}
Here the Wilson line $U_P (x^{\prime },x)=P\exp
(ig\int_x^{x^{\prime }}dy^\rho t^e A_\rho^e (y))$ is introduced in
order that the operator recovers the locally gauge invariance as
pointed out in the QED case given in Sec.III. Note that collecting
Eqs.(36)-(39) gives the expression of $\int d^4x\delta_T L_{QCD}$
which is purely kinematical. To take into account the dynamics of
the system\cite{he1} and also to simplify the expression of $\int
d^4x\delta_T L_{QCD}$, we use the QCD equation of motion for gluon
fields
\begin{equation}
\partial ^\lambda F^a_{\lambda\mu}+gf^{abc}F^b_{\mu\lambda}A^c_{\lambda}
 + g\bar{\psi}\gamma_{\mu}t^{a}\psi + \frac{1}{\xi}\partial _{\mu} (\partial
^{\lambda}A_{\lambda}^a) + gf^{abc}(\partial _{\mu}\bar{c}^b)c^c =
0.
\end{equation}
 We thus obtain
\begin{eqnarray}
& &\int d^4x\delta_T L_{QCD}\nonumber \\
&=& \int d^4x \frac{1}{2}\omega
\epsilon^{\mu\nu}\{\frac{i}{4}g{\lim _{x^{\prime }\rightarrow
x}}(\partial _\lambda ^x-\partial _\lambda ^{x^{\prime }})
\bar{\psi} (x^{\prime}) S_{\lambda \mu \nu } \{U_P(x^{\prime },x),
t^{a}\} \psi (x)c^a(x)
-mg\bar{\psi}(x)\sigma _{\mu \nu }t^{a}\psi (x)c^{a}(x)\nonumber \\
& & + \frac{1}{2}g\bar{\psi}(x)\gamma_{\mu}
t^{a}\psi(x)D_{\nu}^{ab}c_{b}(x) -
\frac{1}{2}g\bar{\psi}(x)\gamma_{\nu}
t^{a}\psi(x)D_{\mu}^{ab}c_{b}(x)\}.
\end{eqnarray}

 Since the generating functional of QCD and the measure of functional integral
are both invariant under the transverse symmetry transformations
given by Eq.(35), such transformations then lead to following
identity (from the identity similar to Eq.(7)) for the functional
integral over two fermion and one anti-ghost fields:
\begin{equation}
0=\int D[\overline{\psi},\psi,A,c,\bar{c}]e^{i\int d^4xL_{QCD}}\{
i\int d^4x(\delta_T
L_{QCD}(x))\psi(x_1)\overline{\psi}(x_2)\bar{c}^e(0)
+\delta_T(\psi(x_1)\overline{\psi}(x_2)\bar{c}^e(0))\}.
\end{equation}
Substituting Eqs.(35) and (41) into this identity and then taking
the coefficient of $\frac{1}{2}\omega\epsilon^{\mu \nu}$ and
dividing the generating functional $Z[J=0]$, we obtain the
transverse Slavnov-Taylor identity for the quark-gluon vertex in
coordinate space:
\begin{eqnarray}
& &\left\langle 0\left| Tgj_{\mu}^a(x)\psi (x_1)\bar{\psi}(x_2)
(D_{\nu}^{ab}c^{b}(x))\bar{c}^{e}(0)\right| 0\right\rangle -
\left\langle 0\left| Tgj_{\nu}^a(x)\psi (x_1)\bar{\psi}(x_2)
(D_{\mu}^{ab}c^{b}(x))\bar{c}^{e}(0)
\right| 0\right\rangle \nonumber \\
&=& igt^{a}\sigma _{\mu \nu }\left\langle 0\left| T\psi
(x_1)\bar{\psi}(x_2)c^{a}(x)\bar{c}^{e}(0)\right| 0\right\rangle
\delta ^4(x_1-x) +ig\left\langle 0\left| T\psi
(x_1)\bar{\psi}(x_2)c^{a}(x)\bar{c}^{e}(0)\right| 0\right\rangle
\sigma _{\mu \nu }t^{a}\delta ^4(x_2-x)\nonumber \\
& &+2mg\left\langle 0\left| T\bar{\psi}(x)\sigma _{\mu \nu
}t^{a}\psi (x)\psi (x_1)
\bar{\psi}(x_2)c^{a}(x)\bar{c}^{e}(0)\right| 0\right\rangle \nonumber \\
& &+ \frac{i}{2}g{\lim _{x^{\prime }\rightarrow x}}(\partial
_\lambda ^x-\partial _\lambda ^{x^{\prime }}) \varepsilon
_{\lambda \mu \nu \rho }\left\langle 0\left| T\bar{\psi}
(x^{\prime}) \gamma ^\rho \gamma _5\{U_P(x^{\prime },x), t^{a}\}
\psi (x)\psi(x_1)\bar{\psi}(x_2)c^{a}(x)\bar{c}^{e}(0)
\right|0\right\rangle ,
\end{eqnarray}
where $j^{a}_{\mu}(x)=\bar{\psi}(x)\gamma_{\mu} t^{a}\psi (x)$.

Note that each term in the transverse ST identity (43) contains a
disconnected part plus connected terms due to the quark-ghost
scattering. To understand how to make Fourier transformation for
such kind of term, let us first discuss a more simple case -- the
another form of the ST identity for the quark-gluon vertex before
making Fourier transformation for Eq.(43). Considering following
transformations:
\begin{equation}
\delta\psi = ig\omega c_{a}t^{a}\psi, \ \ \ \delta\bar{\psi} =
-ig\bar{\psi}t^{a}\omega c_{a},\ \ \ \delta A^{a}_{\mu} = \delta
c^{a}=\delta \bar{c}^{a} = 0,
\end{equation}
and using the procedure similar to the derivations of the WT
identity (9) and the transverse WT identity (16), we can obtain an
expression of the ST identity for the quark-gluon vertex in
coordinate space:
\begin{eqnarray}
& &\left\langle 0\left| Tgj_{\mu}^a(x)\psi (x_1)\bar{\psi}(x_2)
(D^{\mu}_{ab}c^{b}(x))\bar{c}^{e}(0)\right| 0\right\rangle  \nonumber \\
&=& gt^{a}\left\langle 0\left| T\psi
(x_1)\bar{\psi}(x_2)c^{a}(x)\bar{c}^{e}(0)\right| 0\right\rangle
\delta ^4(x-x_1) -g\left\langle 0\left| T\psi
(x_1)\bar{\psi}(x_2)t^{a}c^{a}(x)\bar{c}^{e}(0)\right|
0\right\rangle t^{a}\delta(x-x_2),
\end{eqnarray}
which should be equivalent to the ST identity (30). Hence one can
give the relation between the quark-gluon vertex defined from
$\langle 0|T \psi(x_{1}) \bar{\psi}(x_{2})A^a_{\mu}(x) |0\rangle $
and that from $\langle 0|T j^a_{\mu}(x)\psi(x_{1})
\bar{\psi}(x_{2})|0\rangle $ by the following discussion.

Notice that each term in Eq.(45) contains a disconnected part plus
terms due to the quark-ghost scattering. For instance, the first
term
\begin{eqnarray}
& &\left\langle 0\left| Tj_{\mu}^a(x)\psi (x_1)\bar{\psi}(x_2)
 D^{\mu}_{ab}c^{b}(x)\bar{c}^{e}(0)\right| 0\right\rangle  \nonumber \\
 &=& \left\langle 0\left| Tj_{\mu}^a(x)\psi
(x_1)\bar{\psi}(x_2)\right| 0\right\rangle \left\langle
0\left|TD^{\mu}_{ab}c^{b}(x)\bar{c}^{e}(0)\right| 0\right\rangle +
connected .
\end{eqnarray}
 Then the part $\left\langle 0\left| Tj_{\mu}^a
(x)\psi (x_1)\bar{\psi}(x_2)\right| 0\right\rangle $ can be
decomposed into the three-point proper vertex and the quark
propagator as given by Eq.(11) in the Abelian case. The
contribution of the part $\left\langle
0\left|TD_{\mu}^{ab}c^{b}(x)\bar{c}^{e}(0)\right| 0\right\rangle$
is given by using the identity
\begin{equation}
\int d^4xe^{-iq\cdot x}\left\langle
0\left|TD_{\mu}^{ab}c^{b}(x)\bar{c}^e(0)\right| 0\right\rangle =
\delta ^{ae}q_{\mu}/q^2.
\end{equation}
 Thus Fourier transforming the first term of Eq.(45) leads to $iS_F(p_1)\tilde{\Gamma}
_V^{a\mu}(p_1,p_2)(1-B^{(\mu)}_{(D)6})iS_F(p_2)q^{\nu}/q^2$, where
$B^{(\mu)}_{(D)6}$ is the relative 6-point(body) quark-ghost
scattering kernel from the connected term in Eq.(46) and
$\tilde{\Gamma} _V^{a \mu}$ is the vector vertex defined from
$\left\langle 0\left| Tj_{\mu}^a(x)\psi
(x_1)\bar{\psi}(x_2)\right| 0\right\rangle$ as given in the
Abelian case \cite{he1,he2,he4}. Hence Fourier transforming
Eq.(45) obtains following expression of the ST identity for the
quark-gluon vertex in momentum space:
\begin{equation}
q_\mu\tilde{\Gamma}^{a\mu}_{V}(p_1,p_2;q)(1-B^{(\mu)}_{(D)6}(p_1,p_2))
= [ S_F^{-1}(p_1)(t^{a}-B^{a}_4(p_1,p_2)) -
(t^{a}-B^{a}_4(p_1,p_2))S_F^{-1}(p_2) ]G(q^2).
\end{equation}
Because the ST identity (48) is equivalent to Eq.(32), we thus
obtain the relation
\begin{equation}
\Gamma _V^{a \mu}=\tilde{\Gamma} _V^{a \mu}(1-B^{(\mu)}_{(D)6}).
\end{equation}
On the other hand, integrating the term involving
$\partial_{\mu}^xc^a(x)$  by part and using the QCD equation of
motion for quark field: $(i\gamma^{\mu}D_{\mu} - m)\psi = 0$, we
have
\begin{eqnarray}
& &\int d^4x_1d^4x_2d^4xe^{ip_1\cdot x_1-ip_2\cdot x_2-iq\cdot
x}\left\langle 0\left| Tj_{\mu}^a(x)\psi (x_1)\bar{\psi}(x_2)
 D^{\mu}_{ab}c^{b}(x)\bar{c}^{e}(0)\right| 0\right\rangle \nonumber \\
& &= iq_{\mu}\int d^4x_1d^4x_2d^4xe^{ip_1\cdot x_1-ip_2\cdot
x_2-iq\cdot x}\left\langle 0\left| Tj_{\mu}^a(x)\psi
(x_1)\bar{\psi}(x_2)
 c^{a}(x)\bar{c}^{e}(0)\right| 0\right\rangle \nonumber \\
& &= (2 \pi)^{4}
\delta^{4}(p_{1}-p_{2}-q)iS_F(p_1)q_{\mu}\tilde{\Gamma} _V^{a
 \mu}(p_1,p_2)(1-B^{(\mu)}_{6})iS_F(p_2)G(q^2)/q^2,
\end{eqnarray}
where $B_6^{(\mu)}$ is the 6-body quark-ghost scattering kernel
from the connected term in $\left\langle 0\left| Tj_{\mu}^a(x)\psi
(x_1)\bar{\psi}(x_2)c^{a}(x)\bar{c}^{e}(0)\right| 0\right\rangle$.
We thus find
\begin{eqnarray}
\Gamma _V^{a \mu}(p_1,p_2)&=&\tilde{\Gamma} _V^{a
\mu}(p_1,p_2)(1-B^{(\mu)}_{(D)6}(p_1,p_2)) \nonumber \\
&=&\tilde{\Gamma} _V^{a
 \mu}(p_1,p_2)(1-B^{(\mu)}_{6}(p_1,p_2))G(q^2),
\end{eqnarray}
where $\Gamma _V^{a \mu}$ is defined by Eq.(31), while
$\tilde{\Gamma} _V^{a \mu}$ is defined by a similar equation as
given by Eq.(11). The axial-vector vertex and the tensor vertex
also satisfy similar relations.

Using these relations and by the similar procedure for deriving
the ST identity (32) from Eq.(30) and for deriving the ST identity
(48) from Eq.(45), we can perform the Fourier transformation for
the identity (43), which leads to the transverse Slavnov-Taylor
identity for the quark-gluon vertex in momentum space:
\begin{eqnarray}
& &iq^\mu \Gamma _V^{a \nu}(p_1,p_2)-
iq^\nu \Gamma _V^{a \mu}(p_1,p_2)\nonumber \\
&=&[S_F^{-1}(p_1)\sigma ^{\mu \nu }(t^{a}-B^{a}_4(p_1,p_2))
+(t^{a}-B^{a}_4(p_1,p_2))\sigma^{\mu \nu }S_F^{-1}(p_2)]G(q^2)\nonumber \\
&+& 2m\Gamma _T^{a\mu \nu }(p_1,p_2) + (p_{1\lambda }+p_{2\lambda
})\varepsilon ^{\lambda \mu \nu \rho }\Gamma ^{a}_{A\rho
}(p_1,p_2) -\int \frac{d^{4}k}{(2 \pi)^{4}}2k_{\lambda}
\varepsilon ^{\lambda \mu \nu \rho }\Gamma ^{a}_{A\rho
}(p_1,p_2;k).
\end{eqnarray}
Here $\Gamma _T^{a\mu \nu }=\tilde{\Gamma} _T^{a\mu \nu
}(1-B^{(\mu\nu)}_6)G(q^2)$, $\Gamma ^{a}_{A\rho }= \tilde{\Gamma}
^{a}_{A\rho }(1-B^{(\rho 5)}_{(D)6})=\tilde{\Gamma} ^{a}_{A\rho
}(1-B^{(\rho 5)}_{6})G(q^2)$ and $\Gamma ^{a}_{A\rho }(p_1,p_2;k)=
\tilde{\Gamma} ^{a}_{A\rho }(p_1,p_2;k)(1-B^{(\rho
5)}_6(p_1,p_2;k))G(q^2)$, where $B^{(\mu\nu)}_6$ and $B^{(\rho
5)}_6$ are the 6-body quark-ghost scattering kernels from the
relative connected terms, $\tilde{\Gamma} _T^{a\mu \nu }$ and
$\tilde{\Gamma} ^{a}_{A\rho }$ are respectively the tensor and
axial-vector vertices defined as that in the Abelian case
\cite{he1,he2}. The four-point-like non-local axial-vector vertex
$\Gamma^a_{A\rho}(p_{1}, p_{2};k)$ is defined by the Fourier
transformation of the last matrix element in Eq.(43):
\begin{eqnarray}
& &\int d^{4}x d^{4}x^{\prime}d^{4}x_{1} d^{4}x_{2}
e^{i(p_{1}\cdot x_{1} - p_{2}\cdot  x_{2} + (p_2-k)\cdot x -
(p_1-k)\cdot x^{\prime})} \langle 0|T \bar{\psi}(x^{\prime
})\gamma _\rho \gamma _5 \{U_P(x^{\prime },x), t^{b}\}
\psi (x)\psi(x_1)\bar{\psi}(x_2)c^{b}(x)\bar{c}^{a}(0) |0\rangle \nonumber \\
&=& (2 \pi)^{4} \delta^{4}(p_{1} - p_{2} - q) iS_{F}(p_{1})
\Gamma^a_{A\rho}(p_{1}, p_{2};k) iS_{F}(p_{2}) .
\end{eqnarray}

\subsection{Transverse axial-vector Slavnov-Taylor identity}

The transverse ST identity (52) shows that the transverse part of
the quark-gluon vertex is related to the tensor and axial-vector
vertices. In the case of $m=0$, the contribution of $\Gamma
_T^{a\mu \nu }$ disappears. Therefore, in this case to constrain
completely the quark-gluon vertex, the constraint relation for the
transverse part of $\Gamma ^{a}_{A\rho }$ ( the longitudinal part
of $\Gamma ^{a}_{A\rho }$ does not contribute to Eq.(52) due to
the factor $\varepsilon ^{\lambda \mu \nu \rho }$ ) is required to
build as well. This can be performed by using the transverse
symmetry transformations associated with the chiral
transformations of the fields in QCD.

The infinitesimal chiral transformations in QCD can be written as
\begin{equation}
\delta^{(5)}\psi = ig\omega c_{a}t^{a}\gamma_5\psi, \ \ \
\delta^{(5)}\bar{\psi} = ig\bar{\psi}\gamma_5t^{a}\omega c_{a}, \
\ \ \delta^{(5)} A^{a}_{\mu} = 0,\ \ \ \delta^{(5)} c^{a} =
\delta^{(5)} \bar{c}^{a} = 0.
\end{equation}
By using definition (3), we obtain the infinitesimal transverse
chiral- transformations associated with Eq.(54):
\begin{equation}
\delta^{(5)}_{T}\psi = \frac{1}{4}g \epsilon^{\mu\nu}\omega
c_{a}t^{a}\sigma_{\mu\nu}\gamma_5 \psi, \ \ \
\delta^{(5)}_{T}\bar{\psi} = -
\frac{1}{4}g\epsilon^{\mu\nu}\bar{\psi}\sigma_{\mu\nu}\gamma_5
t^{a}\omega c_{a},\ \ \ \delta^{(5)}_{T} A^{a}_{\mu} =
\delta^{(5)}_{T} c^{a} = \delta^{(5)}_{T} \bar{c}^{a} = 0.
\end{equation}

Such transverse chiral-transformations lead to $\int d^4xL_{QCD}
\longrightarrow \int d^4xL_{QCD} + \int d^4x\delta^{(5)}_T
L_{QCD}$, where
\begin{eqnarray}
& &\int d^4x\delta_T^{(5)} L_{QCD}\nonumber \\
&=& \int d^4x \frac{1}{4}\omega g
\epsilon^{\mu\nu}\{\frac{i}{2}{\lim _{x^{\prime }\rightarrow
x}}(\partial _\lambda ^x-\partial _\lambda ^{x^{\prime }})
\bar{\psi} (x^{\prime}) S^{\lambda \mu \nu } \{U_P(x^{\prime },x),
t^{a}\} \gamma_5 \psi (x)c^a(x) \nonumber \\
& & + j^a_{5\mu}(x)D_{\nu}^{ab}c_{b}(x) -
j^a_{5\nu}(x)D_{\mu}^{ab}c_{b}(x)\}
\end{eqnarray}
with $j^a_{5\mu}=\bar{\psi}(x)\gamma_{\mu}\gamma_5 t^{a}\psi(x)$.

Note that there is no transverse axial-anomaly of QCD, since the
axial-anomaly of QCD should be described by the Abelian result,
supplemented by an appropriate group theory factor, and there is
no transverse axial-anomaly in the Abelian case\cite{he6,sun}.
Correspondingly, the measure of functional integration is
invariant under such transverse chiral transformations, which can
be checked by the explicit calculation. As a result, the gauge
invariance of the generating functional of QCD together with the
invariance of the measure of functional integration in such
transverse chiral-transformations lead to following identity
\begin{equation}
0=\int D[\overline{\psi},\psi,A,c,\bar{c}]e^{i\int d^4xL_{QCD}}\{
i\int d^4x(\delta_T^{(5)}
L_{QCD})\psi(x_1)\overline{\psi}(x_2)\bar{c}^e(0)
+\delta_T^{(5)}(\psi(x_1)\overline{\psi}(x_2)\bar{c}^e(0))\}.
\end{equation}
Substituting Eqs.(55) and (56) into this identity and then taking
the coefficient of $\frac{1}{2}\omega\epsilon^{\mu \nu}$ and
dividing the generating functional $Z[J=0]$, we obtain the
transverse ST identity for the axial-vector quark-gluon vertex in
coordinate space:
\begin{eqnarray}
& &\left\langle 0\left| Tj_{5\mu}^a(x)\psi (x_1)\bar{\psi}(x_2)
(D_{\nu}^{ab}c^{b}(x))\bar{c}^{e}(0)\right| 0\right\rangle -
\left\langle 0\left| Tj_{5\nu}^a(x)\psi (x_1)\bar{\psi}(x_2)
(D_{\mu}^{ab}c^{b}(x))\bar{c}^{e}(0)
\right| 0\right\rangle \nonumber \\
&=& i\sigma _{\mu \nu }\gamma_5\left\langle 0\left| Tt^{a}\psi
(x_1)\bar{\psi}(x_2)c^{a}(x)\bar{c}^{e}(0)\right| 0\right\rangle
\delta ^4(x_1-x) -i\left\langle 0\left| T\psi
(x_1)\bar{\psi}(x_2)t^{a}c^{a}(x)\bar{c}^{e}(0)\right|
0\right\rangle
\sigma _{\mu \nu }\gamma_5\delta ^4(x_2-x)\nonumber \\
& &+ \frac{i}{2}{\lim _{x^{\prime }\rightarrow x}}(\partial
_\lambda ^x-\partial _\lambda ^{x^{\prime }}) \varepsilon
_{\lambda \mu \nu \rho }\left\langle 0\left| T\bar{\psi}
(x^{\prime}) \gamma ^\rho \{U_P(x^{\prime },x), t^{a}\} \psi
(x)\psi(x_1)\bar{\psi}(x_2)c^{a}(x)\bar{c}^{e}(0)
\right|0\right\rangle .
\end{eqnarray}

Using the similar procedure for obtaining Eq.(52) from Eq.(43),
Fourier transforming Eq.(58) leads to the transverse axial-vector
Slavnov-Taylor identity in momentum space:
\begin{eqnarray}
& &iq^\mu \Gamma _A^{a \nu}(p_1,p_2)-
iq^\nu \Gamma _A^{a \mu}(p_1,p_2)\nonumber \\
&=&[S_F^{-1}(p_1)\sigma ^{\mu \nu}\gamma^5(t^{a}-B^{a}_4(p_1,p_2))
-(t^{a}-B^{a}_4(p_1,p_2))\sigma^{\mu \nu }\gamma^5S_F^{-1}(p_2)]G(q^2)\nonumber \\
&+&(p_{1\lambda }+p_{2\lambda })\varepsilon ^{\lambda \mu \nu \rho
}\Gamma ^{a}_{V\rho }(p_1,p_2) -\int \frac{d^{4}k}{(2
\pi)^{4}}2k_{\lambda} \varepsilon ^{\lambda \mu \nu \rho }\Gamma
^{a}_{V\rho }(p_1,p_2;k),
\end{eqnarray}
where the notations are same as that given in last subsection, and
 $\Gamma ^a_{V\rho}(p_1,p_2;k)$ is defined by the Fourier
transformation of the last matrix element in Eq.(58), which is
given from Eq.(53) by replacing $\gamma _\rho\gamma _5$ and
$\Gamma^a_{A\rho}$ with $\gamma _\rho$ and $\Gamma^a_{V\rho}$,
respectively.

The transverse Slavnov-Taylor identities (52) and (59) are the
primary results derived in terms of the transverse symmetry
transformations associated with the BRST symmetry and the chiral
transformations in QCD. They together with the ST identity (32)
form a complete set of Slavnov-Taylor relations for the
quark-gluon vertex in the case of massless fermion.

\section{The Quark-Gluon Vertex Function in QCD}

Now let us derive the the quark-gluon vertex function $\Gamma
_V^{a\mu} $ by consistently solving this set of ST relations for
the vector and the axial-vector quark-gluon vertex functions in
the case of massless fermion. To do this, multiplying both sides
of Eqs.(52) and (59) by $iq_\nu $, and then moving the terms
proportional to $q_\nu \Gamma _V^\nu $ and $q_\nu \Gamma _A^\nu $
into the right-hand side of the equations, we thus have
\begin{eqnarray}
q^2\Gamma _V^{a\mu} (p_1,p_2) &=&q^\mu [q_\nu \Gamma _V^{a\nu}
(p_1,p_2)]+ iq_\nu [S_F^{-1}(p_1)\sigma ^{\mu \nu }(t^a-B^a_4) +
(t^a-B^a_4)\sigma ^{\mu \nu }S_F^{-1}(p_2)]G(q^2)
\nonumber \\
& &+i(p_{1\lambda }+ p_{2\lambda })q_\nu \varepsilon ^{\lambda \mu
\nu \rho }\Gamma ^a_{A\rho }(p_1,p_2) - iq_{\nu}C_A^{a\mu\nu},
\end{eqnarray}
\begin{eqnarray}
q^2\Gamma _A^{a\mu} (p_1,p_2) &=&q^\mu [q_\nu \Gamma _A^{a\nu}
(p_1,p_2)]+ iq_\nu [S_F^{-1}(p_1)\sigma ^{\mu \nu }\gamma
_5(t^a-B^a_4)
- (t^a-B^a_4)\sigma ^{\mu \nu }\gamma _5S_F^{-1}(p_2)]G(q^2)\nonumber \\
& &+i(p_{1\lambda }+p_{2\lambda })q_\nu \varepsilon ^{\lambda \mu
\nu \rho }\Gamma ^a_{V\rho }(p_1,p_2) - iq_{\nu}C_V^{a\mu\nu},
\end{eqnarray}
where
\begin{equation}
C^{a\mu\nu}_A = \int \frac{d^{4}k}{(2 \pi)^{4}}2k_{\lambda}
\varepsilon ^{\lambda \mu \nu \rho }\Gamma ^a_{A\rho }(p_1,p_2;k),
\end{equation}
\begin{equation}
C^{a\mu\nu}_V = \int \frac{d^{4}k}{(2 \pi)^{4}}2k_{\lambda}
\varepsilon ^{\lambda \mu \nu \rho }\Gamma ^a_{V\rho }(p_1,p_2;k),
\end{equation}
which are non-local vertex terms. Substituting Eq.(61) into
Eq.(60) and using the ST identity (32) and following identities
\begin{eqnarray}
& &q_{\nu}q_{\alpha}(p_{1\lambda }+p_{2\lambda }) \varepsilon
^{\lambda \mu \nu \rho } \sigma^{\rho\alpha}\gamma _5
\nonumber \\
&=&i[q_{\nu}q\cdot(p_{1}+p_{2})\sigma ^{\mu \nu} - q^2
(p_{1\lambda }+p_{2\lambda })\sigma ^{\mu \lambda} -
q^{\mu}q_{\nu}(p_{1\lambda }+p_{2\lambda })\sigma ^{\lambda\nu}]
\end{eqnarray}
and
\begin{eqnarray}
& &q_{\nu}q_\alpha(p_{1\lambda }+p_{2\lambda })
 (p_{1\beta }+p_{2\beta })\varepsilon ^{\lambda \mu \nu \rho }
\varepsilon ^{\beta \rho \alpha \delta }\Gamma^a _{V\delta}\nonumber \\
&=&[q^2(p_1 + p_2)^2 - ((p_1 + p_2)\cdot q)^2 ]\Gamma _{V}^{a\mu}
+ [(p_1 + p_2)\cdot q q^{\mu} - q^2(p_1^\mu +p_2^\mu )] (p_{1\nu}
+p_{2\nu} )\Gamma _{V}^{a\nu}\nonumber \\
& & +[(p_1 + p_2)\cdot q(p_1^\mu +p_2^\mu ) - (p_1 + p_2)^2
q^{\mu}]q_\nu \Gamma _V^{a\nu},
\end{eqnarray}
 after self-consistent iterating, we finally obtain the
 quark-gluon vertex function ($m=0$ case) of involving both the longitudinal part
 of the vertex, $\Gamma _{V(L)}^{a\mu}$, and the transverse part of the vertex,
 $\Gamma _{V(T)}^{a\mu}$ :
\begin{equation}
\Gamma _V^{a\mu} (p_1,p_2)=\Gamma _{V(L)}^{a\mu} (p_1,p_2) +\Gamma
_{V(T)}^{a\mu} (p_1,p_2),
\end{equation}
\begin{eqnarray}
& &\Gamma _{V(L)}^{a\mu} (p_1,p_2) = q^{-2}q^\mu [q_\nu \Gamma
_V^{a\nu}(p_1,p_2)]\nonumber \\
&=&q^\mu[S_F^{-1}(p_1)(t^{a}-B^{a}_4(p_1,p_2))-(t^{a}-B^{a}_4(p_1,p_2))S_F^{-1}(p_2)
]G(q^2)/q^2,
\end{eqnarray}
\begin{eqnarray}
 & &\Gamma_{V(T)}^{a\mu} (p_1,p_2)= q^{-2}iq_{\nu}[ iq^{\mu}\Gamma
_V^{a\nu}(p_1,p_2)-iq^{\nu}\Gamma _V^{a\mu}(p_1,p_2)]\nonumber \\
&=&[ q^2+(p_1+p_2)^2-((p_1+p_2)\cdot q)^2q^{-2}]^{-1}G(q^2)/q^2\nonumber \\
& &\times \{
 i[S_F^{-1}(p_1)\sigma^{\mu\nu}(t^{a}-B^{a}_4(p_1,p_2))
+(t^{a}-B^{a}_4(p_1,p_2))\sigma ^{\mu \nu}S_F^{-1}(p_2)]q_\nu q^2\nonumber \\
& &+i[S_F^{-1}(p_1)\sigma ^{\mu \lambda}(t^{a}-B^{a}_4(p_1,p_2))
-(t^{a}-B^{a}_4(p_1,p_2))\sigma ^{\mu \lambda }S_F^{-1}(p_2)]
(p_{1\lambda }+p_{2\lambda })q^2\nonumber \\
& &+i[S_F^{-1}(p_1)\sigma ^{\lambda \nu }(t^{a}-B^{a}_4(p_1,p_2))
-(t^{a}-B^{a}_4(p_1,p_2))\sigma ^{\lambda \nu }S_F^{-1}(p_2)]
q_\nu (p_{1\lambda }+p_{2\lambda })q^\mu \nonumber \\
& &-i[S_F^{-1}(p_1)\sigma ^{\mu \nu }(t^{a}-B^{a}_4(p_1,p_2))
-(t^{a}-B^{a}_4(p_1,p_2))\sigma ^{\mu \nu }S_F^{-1}(p_2)]
q_\nu (p_1+p_2)\cdot q \nonumber \\
& &+ i[S_F^{-1}(p_1)\sigma ^{\lambda \nu}(t^{a}-B^{a}_4(p_1,p_2))
+(t^{a}-B^{a}_4(p_1,p_2))\sigma ^{\lambda \nu }S_F^{-1}(p_2)]\nonumber \\
& &\times q_\nu (p_{1\lambda }+p_{2\lambda })
[p_1^{\mu}+p_2^{\mu} - (p_1+p_2)\cdot q q^\mu q^{-2}] \nonumber \\
& & -iq_\nu q^2 \bar{C}^{a\mu\nu}_A(p_1,p_2) +
q_{\nu}q_{\alpha}(p_{1\lambda }+p_{2\lambda }) \varepsilon
^{\lambda \mu \nu \beta }\bar{C}^{a\beta\alpha}_V(p_1,p_2) \nonumber \\
& & -iq_\nu (p_{1\lambda }+p_{2\lambda })[p_1^{\mu}+p_2^{\mu} -
(p_1+p_2)\cdot q q^\mu q^{-2}]\bar{C}^{a\lambda\nu}_A(p_1,p_2)\},
\end{eqnarray}
where
$\bar{C}^{a\mu\nu}_A(p_1,p_2)=C^{a\mu\nu}_A(p_1,p_2)G^{-1}(q^2)
=\int \frac{d^{4}k}{(2 \pi)^{4}}2k_{\lambda} \varepsilon ^{\lambda
\mu \nu \rho }\tilde{\Gamma} ^{a}_{A\rho }(p_1,p_2;k)(1-B^{(\rho
5)}_6(p_1,p_2;k))$ and $\bar{C}^{a\beta\alpha}_V(p_1,p_2) =
C^{a\beta\alpha}_V(p_1,p_2)G^{-1}(q^2)=\int \frac{d^{4}k}{(2
\pi)^{4}}2k_{\lambda} \varepsilon ^{\lambda \beta\alpha \rho
}\tilde{\Gamma} ^{a}_{V\rho }(p_1,p_2;k)(1-B^{(\rho
)}_6(p_1,p_2;k))$. Eqs.(66)-(68) describe the constraints on the
structure of the quark-gluon vertex function($m=0$ case) imposed
from the gauge symmetry alone of QCD, showing how the quark-gluon
vertex function relates to the quark propagator, the ghost
dressing function, the quark-ghost scattering kernels and the
four-point-like non-local vertex terms. The axial-vector
quark-gluon vertex function and the quark-gluon vertex function
with fermion mass ($m\not=0$) can be similarly derived.

\section{Conclusions and Discussions}

This paper introduces the transverse symmetry transformations
associated with the normal symmetry transformations in gauge
theories, which enables us to build the transverse constraints on
the vertex functions in gauge theories. This has been tested at
first by using such approach to reproduce the transverse WT
relations that obtained already in canonical field theory approach
in QED. Then by using the transverse symmetry transformations
associated with the BRST transformations and chiral
transformations in QCD and in terms of the path-integral approach,
we have derived the transverse ST identities for the vector and
axial-vector quark-gluon vertices, respectively. Based on the set
of normal and transverse ST identities, we have further obtained
an expression of the quark-gluon vertex function.

It is important to emphasize that, while the BRST symmetry leads
to the ST identity which constrains the longitudinal part of the
quark-gluon vertex from the gauge symmetry, the transverse
symmetry transformations associated with the BRST and the chiral
transformations lead respectively to the transverse ST identities
for the vector and the axial-vector quark-gluon vertices, which
have the potential to constrain the transverse part of the
quark-gluon vertex from the gauge symmetry, and hence the
expression of the quark-gluon vertex function given by
Eqs.(66)-(68) describes the constraints on the quark-gluon vertex
structure imposed from the gauge symmetry alone of QCD theory in
the massless case. Hence, such a quark-gluon vertex function
should be satisfied both perturbatively and nonperturbatively and
then has the potential to unravel the non-Abelian structure of the
quark-gluon vertex. In these aspects, some comments are given as
follows.

At first, it can be checked that the transverse ST identities (52)
and (59) and then the quark-gluon vertex function given by
Eqs.(66)-(68) should hold in perturbation theory by performing the
corresponding one-loop calculations as done in the Abelian theory
QED case\cite{he4, penn, he5, he7}. As shown in the Abelian QED
case, the non-local vertex terms in the transverse WT identities
are essential to insure that the fermion-boson vertex derived
based on the set of normal and transverse WT identities holds to
one-loop order, and are responsible for multiplicative
renormalizability in perturbation theory. The situation should be
similar for non-Abelian QCD case. Besides, the quark-ghost
scattering kernels are responsible for the one-loop non-Abelian
vertex diagram as shown by the one-loop calculations of 4-point
quark-ghost scattering kernel $B^a_4$ \cite{davy}. These tedious
one-loop calculations for these transverse ST identities and the
quark-gluon vertex given by Eqs.(66)-(68) will remain to be
performed in the further work.

Second, if the 4-point and 6-point quark-ghost scattering kernels
are neglected, the quark-gluon vertex function will reduce to the
Abelian-type vertex function\cite{he7} multiplying the ghost
dressing function. Lattice QCD calculations\cite{skul} for the
quark-gluon vertex in Landau gauge at two specific kinematic
limits ('asymmetric' and 'symmetric') have found substantial
deviations from the Abelian form -- which cannot be described by a
universal function multiplying the Abelian form with the
longitudinal vertex of the Ball-Chiu construction \cite{ball} and
the transverse vertex of the Curtis-Pennington construction
\cite{curt} as given in \cite{fisc}\cite{he8}. Furthermore, recent
lattice data lead to an essentially constant ghost dressing
function in the infrared limit \cite{cucc}. These lattice results,
together with the one-loop calculations, show that the quark-ghost
scattering kernels involved in the quark-gluon vertex must be
non-trivial and are essential for characterizing the non-Abelian
property of the quark-gluon vertex. How to extract the non-trivial
information encoded in these quark-ghost scattering kernels is
required to be studied further.

Third, the longitudinal and transverse parts of the quark-gluon
vertex given respectively by Eqs.(67) and (68) contain several
kinematic singularities at $q^2=0$, which arise from following
reasons: (i) The vertex has been separated into longitudinal and
transverse parts by such a way: $\Gamma _V^{a\mu}(p_1,p_2) =
q^{-2}q^{\mu} [q_{\nu} \Gamma _V^{a\nu}(p_1,p_2)] +
q^{-2}iq_{\nu}[ iq^{\mu}\Gamma _V^{a\nu}(p_1,p_2)-iq^{\nu}\Gamma
_V^{a\mu}(p_1,p_2)]$. In the case that the quark-gluon vertex is
used into the quark propagator DSE, this type of factor $q^{-2}$
might be attributed to the gluon propagator as shown in the
introduction by the expression of the kernel of quark DSE. (ii)
The itinerant procedure performed by substituting the expression
of the axial-vector vertex, Eq.(61), into the expression of the
vector vertex, Eq.(60), leads to the appearance of the factor
$(p_1+p_2)\cdot q q^{-2}$ in the transverse part of the vertex.
The same situation also appears in the Abelian QED case where the
fermion-boson vertex has been expressed in terms of the normal and
transverse WT identities for the vector and axial-vector
vertices\cite{he7}. As shown by the explicit calculations, such a
fermion-boson vertex to one-loop order leads to the same result as
one given in QED perturbation theory\cite{he4,penn}, which does
not exhibit particle-like singularity at $q^2=0$. This result
implies that such a fermion-boson vertex should not exhibit the
particle-like singularity at $q^2=0$\cite{munc,ball}, and hence in
the practical application these kinematic singularities contained
in such a fermion-boson vertex should be cancelled by a proper
procedure like the Ball-Chiu construction\cite{ball} as discussed
by Pennington and Williams for the Abelian QED case\cite{penn}.
The discussion for non-Abelian case should be analogous.

Finally, let us mention the possible application of the transverse
ST identities (52) and (59) and the quark-gluon vertex function
given by Eqs.(66)-(68) in the study of the Dyson-Schwinger
equation (DSE) for the quark propagator. As shown already in the
introduction, the kernel $D_{\mu\nu}(q)\Gamma^{a\nu}(p_1,p_2)$ of
the quark DSE can be naturally separated into the contributions
from transverse and longitudinal parts of the quark-gluon vertex,
which provides important information: Generally, the transverse
and longitudinal parts of the quark-gluon vertex both are
essential for the quark DSE; However, in the Landau gauge QCD with
$\xi=0$ the contribution from longitudinal part of the vertex to
the quark DSE will disappear and then the transverse part of the
vertex will dominate the quark DSE. As a consequence, the
transverse ST identities, which impose the constraints on the
transverse part of the quark-gluon vertex from the gauge symmetry,
will play the crucial role in the study of the DSE for the quark
propagator in Landau gauge QCD. Present work provides the
transverse ST identities (52) and (59) for the vector and
axial-vector quark-gluon vertices and the transverse part of the
quark-gluon vertex function (68) derived from the symmetry
relations of QCD. They together with the ST identity for the
quark-gluon vertex, Eq.(32), provide the bases of an Ansatz for
constructing the quark-gluon vertex being free of kinematic
singularity, like that the Ward-Takahashi identity is a base of
Ansatz for the Ball-Chiu construction \cite{ball} of the Abelian
vertex in QED, for the practical application in the study of the
DSE for the quark propagator. This interesting subject is beyond
the scope of the present work and calls for the further study.

\section*{Acknowledgments}

This work was supported in part by the National Natural Science
Foundation of China under grant No.90303006.


\begin{thebibliography}{99}

\bibitem{ward} J.Ward, Phys.Rev.{\bf 78},182(1950);
         Y.Takahashi, Nuovo Cimento {\bf 6},370(1957).
\bibitem{slav} A.A.Slavnov, Theor. and Math.Phys.{\bf 10},99(1972);
         J.C.Taylor, Nucl.Phys.{\bf B33},436(1971).
\bibitem{robe} C.D.Roberts and A.G.Williams, Prog.Part.Nucl.Phys.{\bf 33}
         477(1994);  R.Alkofer and L.von Smekal, Phys.Rep.{\bf 353}281(2001);
         P.Maris and C.D.Roberts, Int.J.Mod.Phys.{\bf E12},297(2003); and
         references therein.
\bibitem{munc} H.J.Munczk, Phys.Lett.{\bf B175},215(1986);
         C.J.Burden, C.D.Roberts, and A.G.Williams, Phys.Lett.{\bf B285},347(1992);
         F.T.Hawes, P.Maris, and C.D.Roberts, Phys.Lett.{\bf
         B440},353(1998); G. Eichmann, I.C. Cl\"{o}t, R. Alkofer,
         A. Krassnigg, and C.D. Roberts, Phys. Rev. {\bf C 79}, 012202(R) (2009);
         L. Chang, and C.D. Roberts, arXiv: 0903.5461 [nucl-th].
\bibitem{alk1} R.Alkofer, M.Kloker, A.Krossnigg and
        R.F.Wogenbrunn, Phys.Rev.Lett.{\bf 96},022001(2006);
         M.S.Bhagwat and P.C.Tandy, Phys.Rev.{\bf
         D70},094039(2004); A.Bender,W.Detmold,A.W.Thomas and
         C.D.Roberts, Phys.Rev.{\bf C65},065203(2002).
\bibitem{skul} J.Skullerud, P.O.Bowman, A.Kizilersu, D.B.Leinweber
        and A.G.Williams, JHEP{\bf 0304},047(2003).
\bibitem{eich} E.J.Eichten and F.L.Feinberg, Phys.Rev.{\bf
        D10},3254(1974); P.Pascual and R.Tarrach, $QCD$: $Renomalization$
        $for$ $the$ $Practitioner$. Lecture Notes in Physics,V.194(Springer, Berlin,1984).
\bibitem{becc} C.Becchi, A.Rouet, and R.Stora, Phys.Lett.{\bf B52},344(1974);
        Ann.Phys.{\bf 98},287(1976); I.V.Tyutin, Lebedev Institute
        preprint N39(1975, unpublished).
\bibitem{he1} H.X.He, F.C.Khanna and Y.Takahashi, Phys.Lett.{\bf
        B480},222(2000).
\bibitem{he2} H.X.He, Phys.Rev.{\bf C63},025207(2001).
\bibitem{pesk}M.E.Peskin, D.V.Schroeder, An Introduction to
        Quantum Field Theory, Addison-Wesley,1995.
\bibitem{he3} H.X.He and H.W.Yu, Commun.Theor.Phys.(Beijing,China) {\bf 39},
         559(2003).
\bibitem{he4} H.X.He and F.C.Khanna, Int.J.Mod.Phys.{\bf A21},2541(2006).
\bibitem{penn} M.R.Pennington and R.Williams, J.Phys.G:
         Nucl.Part.Phys.{\bf 32},2219(2006).
\bibitem{he5} H.X.He, Int.J.Mod.Phys.{\bf A22},2119(2007).
\bibitem{adle} S.L.Adler, Phys.Rev.{\bf 177},2426(1969); J.S.Bell and
        R.Jackiw, Nuovo Cim.{\bf A60},47(1969).
\bibitem{fuji} K.Fujikawa, Phys.Rev.Lett.{\bf 42},1195(1979).
\bibitem{he6} H.X.He, Phys.Lett.{\bf B507},351(2001).
\bibitem{sun} W.M.Sun, H.S.Zong,X.S.Chen and F.Wang, Phys.Lett.{\bf
         B569},211(2003).
\bibitem{he7} H.X.He, Science in China Series {\bf G51},
         1206(2008); hep-th/0606039.
\bibitem{davy} A.I.Davydychev, P.Osland and L.Saks, Phys.Rev.{\bf
         D63},014022(2000).
\bibitem{ball} J.S.Ball and T.W.Chiu, Phys.Rev.{\bf
         D22},2542(1980).
\bibitem{curt} D.C.Curtis and M.R.Pennington, Phys.Rev.{\bf
         D42},4165(1990).
\bibitem{fisc} C.S.Fischer and R.Alkofer, Phys.Rev.{\bf
         D67},094020(2003).
\bibitem{he8} Note that the B-C type's longitudinal vertex\cite{ball} has been constructed
         in a way free of kinematic singularities by satisfying the Ward
         and WT identities, and the C-P type's transverse
         vertex\cite{curt} has been constructed, by requiring to satisfy
         multiplicative renormalizability, through a nonperturbative
         extension for the asymptotic limit contribution of transverse part
         of one-loop fermion-boson vertex at large fermion momenta
         ($p_1^2 \gg p_2^2$).
\bibitem{cucc} A.Cuccieri, T.Mendes, Infrared behavior and
         infinite-volume limit of gluon and ghost propagators in Yang-Mills
         theories. arXiv:0812.3261(2008).

\end{thebibliography}
\end{document}